\documentclass[pra,twocolumn,a4paper,superscriptaddress,nofootinbib,showpacs]{revtex4}
\usepackage{amsmath}
\usepackage{amsfonts}
\usepackage{color}

\begin{document}
\title{Decoherence-full subsystems and the cryptographic power of a
  private shared reference frame}
\author{Stephen D. Bartlett}
\email{bartlett@physics.uq.edu.au}
\affiliation{School of Physical Sciences, The University of Queensland,
  Queensland 4072, Australia}
\author{Terry Rudolph}
\email{t.rudolph@imperial.ac.uk}
\affiliation{Optics Section, Blackett Laboratory, Imperial College
  London, London SW7 2BZ, United Kingdom}
\author{Robert W. Spekkens}
\email{rspekkens@perimeterinstitute.ca}
\affiliation{Perimeter Institute for Theoretical Physics, 35 King
  St.~N, Waterloo, Ontario N2J 2W9, Canada}
\date{10 September 2004}

\begin{abstract}
  We show that private shared reference frames can be used to perform
  private quantum and private classical communication over a public
  quantum channel.  Such frames constitute a novel type of private
  shared correlation (distinct from private classical keys or shared
  entanglement) useful for cryptography.  We present optimally
  efficient schemes for private quantum and classical communication
  given a finite number of qubits transmitted over an insecure channel
  and given a private shared Cartesian frame and/or a private shared
  reference ordering of the qubits. We show that in this context, it
  is useful to introduce the concept of a \emph{decoherence-full}
  subsystem, wherein every state is mapped to the completely mixed
  state under the action of the decoherence.
\end{abstract}
\maketitle

\section{Introduction}

It is well known that a private classical key can be used for secure
classical communication on a public channel using the Vernam cipher
(one-time pad)~\cite{Ver26}.  Specifically, an $n$-bit string $M$, the
\emph{plain-text}, can be added bit-wise (modulo 2) to a random
$n$-bit string $K$, the \emph{key}, to yield an $n$-bit string $C=M
\oplus K$, the \emph{cipher-text}.  Someone who possesses the key can
retrieve the plain-text from the cipher-text via $M=C \oplus K$;
however, for someone who does not possess the key, $C$ is completely
random and contains no information about $M$.  The cipher-text can
therefore be transmitted over a public channel with complete security.

In quantum cryptography\footnote{Note that we are not here referring
  to quantum key distribution, but rather to the use of a key for
  encoding information.}, quantum rather than classical systems are
used for the transmission (i.e., a quantum cipher-text), allowing for
one or both of the following innovations: (i) the key is quantum,
corresponding to entanglement between the cooperating parties; (ii)
the plain-text is quantum, namely, a quantum state drawn from a set of
states not all of which are orthogonal.

A classical plain-text can be encrypted with a quantum key
(specifically, 2 c-bits can be encrypted using 1 e-bit of entanglement)
by making use of a dense coding protocol~\cite{Ben92}.  A quantum
plain-text can be encrypted with a classical key (specifically, 1
qubit with 2 c-bits) by a scheme known as a private quantum
channel~\cite{Amb00}.  Finally, a quantum plain-text may be encrypted
with a quantum key (1 qubit with 2 e-bits) using the quantum Vernam
cipher~\cite{Leu02} \footnote{Alternative schemes for encrypting 1
  qubit using 2 e-bits are to implement a teleportation protocol for
  the qubit wherein the classical communication is achieved by dense
  coding, or to convert the 2 e-bits into 2 secret c-bits through
  measurement and then use the protocol of~\cite{Amb00}.}. Note that
when the plain-text is quantum, it has been shown that it is possible,
by monitoring for eavesdropping, to recycle the key for future
use~\cite{Leu02,Opp03}.  What all these schemes have in common is that
they make use of \emph{private shared correlations} to encode
information.

In this paper, we wish to consider the applications to
cryptography of a different sort of private shared correlation,
namely, a private \emph{shared reference frame} (SRF).  Two
parties are said to share a reference frame (RF) for some degree
of freedom when there exists an isomorphism between their
experimental operations involving this degree of
freedom~\cite{BRS03a}.  For example, Alice and Bob are said to
share a Cartesian frame, defining an orthogonal trihedron of
spatial orientations, when they can implement the following task.
Alice sends to Bob a spin-1/2 particle aligned along a direction
$\vec{n}$ with respect to her local Cartesian frame. She then
communicates a classical description of this direction to Bob (for
instance, its Euler angles), and Bob must orient his Stern-Gerlach
magnets in such a way that the spin-1/2 particle emerges in the
upper path with certainty.  If Alice and Bob can orient themselves
with respect to the fixed stars, then they will be able to
implement the task above, and thus will be said to share a
Cartesian frame.  An alternative method for sharing a Cartesian
frame is for Alice and Bob to possess, within their respective
labs, sets of gyroscopes that were aligned at a time prior to
Alice and Bob having been separated.

Two parties are said to possess a \emph{private} SRF for some
degree of freedom if the experimental operations of all other
parties fail to be isomorphic to theirs in the sense described
above.  Although it is difficult to imagine how a Cartesian frame
defined by the fixed stars might be made private, it is clear that
if the Cartesian frame is defined by a set of gyroscopes, privacy
amounts to no other party having gyroscopes that are known to be
aligned with those of Alice and Bob.

Unlike either classical or quantum information, which can be
communicated using any degree of freedom one chooses, reference frames
require the transmission of a system with a very specific degree of
freedom~\cite{Per02}.  Two clocks can only be synchronized by the
transmission of physical systems that carry timing information, such
as photons, and two Cartesian frames can only be aligned by the
transmission of physical systems that carry some directional
information, such as spin-1/2 particles.  The optimal way of
establishing a SRF given different sorts of information carriers has
been the subject of many recent
investigations~\cite{GisPop,Per01,Bag01,Per01b,Bag01b,Lin03}.
Recognizing the distinction between SRFs and either classical key or
quantum entanglement has also been important in identifying the
resources that are required for continuous variable teleportation in
quantum optics~\cite{Fur98,Bra99,Rud01,Enk02,Wis02,Wis03,San03,Wis04}.
There have also been several investigations into the impact of
\emph{lacking} the resource of a SRF for various tasks.  These tasks
have included communicating classical and quantum
information~\cite{BRS03a}, accessing entanglement~\cite{Bar03},
discriminating states in a data hiding protocol~\cite{Ver03}, and
implementing successful cheating strategies in two-party cryptographic
protocols such as bit commitment~\cite{KMP03}.

In the present work, we further clarify the nature of SRFs as a
resource, by determining the extent to which \emph{private} SRFs are a
resource for cryptography.

To illustrate the general idea, consider the case where Alice and Bob
share a private Cartesian frame. They can then achieve some private
classical communication as follows: Alice transmits to Bob an
orientable physical system (e.g., a pencil or a gyroscope) after
encoding her message into the relative orientation between this system
and her local reference frame (for instance, by turning her bit string
into a set of Euler angles).  Bob can decrypt the message by measuring
the relative orientation between this system and his local reference
frame.  Because an eavesdropper (Eve) does not have a reference frame
correlated with theirs, she cannot infer any information about the
message from the transmission.

In classical mechanics, it is in principle possible to discriminate
among a continuum of different states of a finite system.  In this
setting, a private shared reference frame together with the
transmission of a finite system would allow for the private
communication of an arbitrarily long message.  However, in quantum
mechanics, finite systems support only a finite number of
distinguishable states, so the question of the private communication
capacity of a private SRF given finite uses of a channel is
non-trivial.  In addition, we can investigate the possibility of
private \emph{quantum} communication.

This paper is structured as follows.  In Sec.~\ref{sec:Example}, we
describe how two parties who share a private Cartesian frame can
privately communicate quantum or classical information using one, two,
or three transmitted spin-1/2 particles.  These examples illustrate
the central concepts of the paper.  In Sec.~\ref{sec:Quantum}, we
present optimally efficient private quantum communication schemes for
arbitrary numbers of transmitted qubits. It is also here that we
properly introduce the concept of a decoherence-full subsystem.  In
Sec.~\ref{sec:Classical}, we present optimally efficient schemes for
private classical communication for large numbers of transmitted
qubits.  Finally, in Sec.~\ref{sec:Conclusions} we conclude with a
discussion of the significance of these results as well as some
directions for future research.

\section{Some simple examples}
\label{sec:Example}

Consider a communication scenario consisting of two parties, a
sender (Alice) and a receiver (Bob), who have access to an insecure
noiseless quantum channel and who possess a private SRF.  Continuing
with our example, we consider spin systems that possess only
rotational degrees of freedom, in which case all local experimental
operations, such as the placement of a Stern-Gerlach magnet, are
performed relative to a local Cartesian frame which is private.

\subsection{One transmitted qubit}

Consider the transmission of a single qubit from Alice to Bob.
As they possess an isomorphism between their experimental operations,
Bob can use the outcomes of his measurements to infer information
about Alice's preparation.  For example, they can communicate a
classical bit by Alice preparing one of an orthogonal pair of states
($|0\rangle$ or $|1\rangle$) and Bob performing the corresponding
projective measurement which reveals the preparation with certainty.

On the other hand, an eavesdropper (Eve) who does not share Alice and
Bob's private SRF cannot correlate the outcomes of her measurements
with Alice's preparations.  To represent the state of the transmitted
qubit, Eve must average over all rotations $\Omega \in$ SU(2) that
could describe the relation between her local RF and theirs.  Thus,
Eve would represent the state of the qubit relative to her
uncorrelated reference frame as
\begin{equation}
  \label{eq:SingleQubitSuperop}
  \mathcal{E}_1(\rho) = \int {\rm d}\Omega\, R(\Omega) \rho R^\dag
  (\Omega) = \tfrac{1}{2}I \, ,
\end{equation}
where $R(\Omega)$ is the spin-1/2 unitary representation of $\Omega
\in$ SU(2), ${\rm d}\Omega$ is the SU(2)-invariant
measure\footnote{The invariant measure is chosen using the maximum
  entropy principle: because Eve has no prior knowledge about Alice's
  RF, she should assume a uniform measure over all possibilities.} and
$I$ is the identity.  Thus, as a result of being uncorrelated with the
private SRF, Eve cannot acquire any information about Alice's
preparation.  Using this single qubit and their private SRF, Alice and
Bob can privately communicate one logical qubit, and thus also one
logical classical bit.

\subsection{Two transmitted qubits:  Decoherence full subspaces}
\label{sec:twotransmittedqubits}

If multiple qubits are transmitted, it is possible for Eve to acquire
some information about the preparation even without access to the
private SRF by performing \emph{relative} measurements on the
qubits~\cite{BRS03b}.  Consider the example of two transmitted qubits,
and suppose that Alice assigns the state $\rho$ to the pair.  Eve does
not know how her RF is oriented relative to Alice's, but she knows
that both qubits were prepared relative to the same RF. Thus, Eve's
description of the pair is obtained from Alice's by averaging over all
rotations $\Omega \in$ SU(2), but with the same rotation applied to
each qubit. Eve therefore describes the pair by the Werner
state~\cite{Wer89}
\begin{align}
  \label{eq:TwoQubitSuperop}
  \mathcal{E}_2(\rho) &= \int {\rm d}\Omega\, R(\Omega)^{\otimes2}
  \rho R^\dag (\Omega)^{\otimes2} \nonumber \\
  &= p_1 (\tfrac{1}{3}\Pi_{j=1}) + p_0 \Pi_{j=0} \, ,
\end{align}
where
\begin{equation}
  p_j = {\rm Tr}(\rho \Pi_{j})\,,
\end{equation}
and where $R(\Omega)^{\otimes2} \equiv R(\Omega) \otimes R(\Omega)$ is the
(reducible) collective representation of SU(2) on two qubits, and
$\Pi_j$ is the projector onto the subspace of total angular momentum
$j$.  It is clear that Eve has some probability of distinguishing
states that differ in the weight they assign to the symmetric ($j=1$)
and antisymmetric ($j=0$) subspaces.  Moreover, she can distinguish
perfectly between the antisymmetric state and a state which lies in
the symmetric subspace.  In other words, despite not sharing the RF,
Eve can still measure the magnitude of the total angular momentum
operator $\hat{J}^2$ and thus acquire information about the
preparation.

Eq.~(\ref{eq:TwoQubitSuperop}) implies that the two-qubit
superoperator $\mathcal{E}_2$ is completely depolarizing on the
three-dimensional symmetric subspace.  In contrast to decoherence-free
subspaces~\cite{Zan97} used in quantum computing, the effect of the
map $\mathcal{E}_2$ on this subspace is irreversible: the
superoperator takes any state on this subspace to a fixed state,
namely, the completely mixed state on this subspace.  In
Section~\ref{sec:Quantum}, we will define subspaces with this property
to be \emph{decoherence-full subspaces}\footnote{Note that the term
  ``decoherence'' has many connotations in the literature.  Here, we
  shall take the term to be synonymous with ``noise'', where this
  noise may arise from ignorance rather than a coupling to the
  environment.}.

By encoding in a decoherence-full subspace, Alice can achieve private
quantum communication.  For instance, Alice can encode a logical
qutrit\footnote{A qutrit is a 3-dimensional generalization of the
  qubit.} state into a state $\rho_S$ of two qubits that has support
entirely within the symmetric subspace.  Bob, sharing the private RF,
can recover this qutrit with perfect fidelity.  However, Eve
identifies all such qutrit states with $\mathcal{E}_2(\rho_S) =
\tfrac{1}{3}\Pi_{j=1}$, the completely mixed state on the $j=1$
subspace, and therefore cannot infer anything about $\rho_S$.  Thus,
using this scheme, a private qutrit can be transmitted from Alice to
Bob using two qubits.

Now consider how many classical bits of information Alice can transmit
privately to Bob. An obvious scheme is for her to encode a classical
trit as three orthogonal states within the symmetric subspace. (For
example, using the three symmetric Bell states $|\psi^+\rangle$,
$|\phi^+\rangle$ and $|\phi^-\rangle$.)  However, this is not the
optimally efficient scheme. Suppose instead that Alice encodes two
classical bits as the four orthogonal states
\begin{equation}
  \label{eq:fourprivatestates}
  \left| i\right\rangle =\frac{1}{2}\left| \psi ^{-}\right\rangle
  +\frac{\sqrt{3}}{2}\left| {\bf n}_{i}\right\rangle \left| {\bf
  n}_{i}\right\rangle \, ,\quad i=1,\ldots,4 \, ,
\end{equation}
where $\left| \psi ^{-}\right\rangle $ is the singlet state and the
$\left| {\bf n}_{i}\right\rangle \left| {\bf n}_{i}\right\rangle $ are
four states in the symmetric subspace with both spins pointed in the
same direction, with the four directions forming a tetrahedron, and
with the phases chosen to ensure orthogonality of the $\left|
  i\right\rangle$~(see~\cite{Mas95}).  It is easy to verify that
\begin{equation}
  \mathcal{E}_{2}(|i\rangle\langle i|)=\tfrac{1}{4}I\,,
\end{equation}
the completely mixed state on the two qubit Hilbert space.  Thus,
these four states are completely distinguishable by Bob but completely
indistinguishable by Eve.  By Holevo's theorem, two classical bits is
the maximum one could possibly communicate by the transmission of two
qubits, so this scheme is optimally efficient.

\subsection{Three transmitted qubits:  Decoherence-full subsystems}

Consider the transmission of three qubits from Alice to Bob. If Alice
prepares these qubits in the state $\rho$, then Eve, who lacks the
SRF, assigns the state
\begin{align}
  \label{eq:ThreeQubitSuperop}
  \mathcal{E}_3(\rho) &= \int {\rm d}\Omega\, R(\Omega)^{\otimes3}
  \rho R^\dag (\Omega)^{\otimes3} \,.
\end{align}
With three qubits, the four-dimensional symmetric subspace consisting
of states with total angular momentum $j=3/2$ is a decoherence-full
subspace: all states on this subspace are mapped by $\mathcal{E}_3$ to
the completely mixed state on this subspace.

The four-dimensional subspace $\mathbb{H}_{j=1/2}$ consisting of
states with total angular momentum $j=1/2$ has a more complex
structure.  This subspace can be given a tensor product structure
(TPS)~\cite{Zan03} as
\begin{equation}
  \label{eq:TwoQubitTPS}
  \mathbb{H}_{j=1/2} = \mathbb{H}_{R} \otimes \mathbb{H}_{P} \, ,
\end{equation}
where $\mathbb{H}_{R}$ is a two-dimensional Hilbert space that carries
the $j=1/2$ irreducible representation of SU(2), and $\mathbb{H}_{P}$
is a two-dimensional Hilbert space that carries the trivial
representation of SU(2).  This TPS does not correspond to the TPS
obtained by combining multiple qubits: it is
\emph{virtual}~\cite{Zan01b}.  We refer to these two factor spaces as
\emph{subsystems}, a concept we will define more precisely in
Section~\ref{sec:Quantum}.  For the moment, we consider how the
superoperator $\mathcal{E}_3$ acts on states in terms of these
subsystems.  Because SU(2) acts irreducibly on $\mathbb{H}_{R}$ and
trivially on $\mathbb{H}_{P}$, the superoperator $\mathcal{E}_3$
restricted to states on $\mathbb{H}_{j=1/2}$ can be expressed as
\begin{equation}
  \label{eq:TwoQubitEonTPS}
  \mathcal{E}_3(\rho_{j=1/2}) = (\mathcal{D}_R \otimes
  \mathcal{I}_P)(\rho_{j=1/2}) \, ,
\end{equation}
where $\mathcal{D}_R$ is the completely depolarizing superoperator on
$\mathbb{H}_{R}$ and $\mathcal{I}_P$ is the identity operation on
$\mathbb{H}_{P}$.  Thus, $\mathcal{E}_3$ takes any product state of
the form $\rho_R \otimes \sigma_P$ to the state $\tfrac{1}{2}I_R
\otimes \sigma_P$.  In fact, $\mathcal{D}_R \otimes \mathcal{I}_P$
maps any state $\rho_{j=1/2}$ on $\mathbb{H}_{R} \otimes
\mathbb{H}_{P}$ to the product state $\tfrac{1}{2}I_R \otimes {\rm
  Tr}_R(\rho_{j=1/2})$, where ${\rm Tr}_R$ is the partial trace over
the subsystem $\mathbb{H}_R$, thus removing all correlations between
the subsystems.  We call the subsystem $\mathbb{H}_{R}$ a
\emph{decoherence-full subsystem}.

We can now express the action of the superoperator $\mathcal{E}_3$ on
an arbitrary state $\rho$ of three qubits as
\begin{equation}
  \label{eq:E3Decomp}
  \mathcal{E}_3(\rho) = p_{3/2} (\tfrac{1}{4}\Pi_{j=3/2}) + p_{1/2}
  (\tfrac{1}{2}\mathbb{I}_R \otimes \rho_P) \, ,
\end{equation}
where
\begin{align}
  p_{j}&= {\rm  Tr}(\rho\Pi_{j})\,, \\
  \rho_P &= \tfrac{1}{p_{1/2}} {\rm Tr}_R(\Pi_{j=1/2}\rho\Pi_{j=1/2}) \, .
\end{align}
Consider the following two options that Alice has for privately
communicating quantum states to Bob using their private SRF: 1)
she can encode quantum states into the decoherence-full $j=3/2$
subspace (allowing private communication of two qubits); 2) she
can encode a qubit state $\rho$ into a product state $\rho\otimes
\sigma_0$ in the $j=1/2$ subspace, where $\sigma_0$ is some fixed
state on $\mathbb{H}_P$. (Using the latter scheme, all states are
represented by Eve as $\tfrac{1}{2}I_R \otimes \sigma_0$, who thus
cannot obtain any information about $\rho$.)  Clearly, using the
$j=3/2$ subspace provides a superior capacity, and we will prove
in Section~\ref{sec:Quantum} that this scheme is optimally
efficient for three qubits.  Note however that for greater numbers
of qubits, the decoherence-full subsystems typically have greater
dimensionality than the decoherence-full subspaces, and schemes
that encode within them are necessary to achieve optimal
efficiency.

For private \emph{classical} communication, the question of optimal
efficiency is much more complex.  One scheme would be for Alice to
encode two c-bits into four orthogonal states within the $j=3/2$
decoherence-full subspace.  Using the $j=1/2$ subspace, it might seem
that the best Alice can do is to encode a single c-bit into two
orthogonal states in the decoherence-full subsystem $\mathbb{H}_R$;
however, there is a better scheme using this subspace.  If Alice
encodes two c-bits into four orthogonal \emph{maximally entangled}
states on the virtual TPS $\mathbb{H}_R \otimes \mathbb{H}_P$, these
states are completely distinguishable by Bob but, using
Eq.~(\ref{eq:E3Decomp}), all map to the same state $\tfrac{1}{2}I_R
\otimes \tfrac{1}{2}I_P$ on $\mathbb{H}_R \otimes \mathbb{H}_P$ under
$\mathcal{E}_3$ and thus are completely indistinguishable from Eve's
perspective.  Thus, using the $j=1/2$ subspace, Alice can privately
transmit two c-bits to Bob, the same number as can be achieved using
the $j=3/2$ subspace.

It turns out that the optimally efficient scheme for private classical
communication uses \emph{both} the $j=3/2$ and $j=1/2$ subspaces.  Let
$|j{=}3/2,\mu\rangle$, $\mu=1,\ldots,4$ be four orthogonal states on
the $j=3/2$ subspace, and let $|j{=}1/2,\mu\rangle$, $\mu=1,\ldots,4$
be four maximally entangled states (as described above) on the $j=1/2$
subspace.  Define the eight orthogonal states
\begin{equation}
  \label{eq:EightOnThreeQubits}
  |b,\mu\rangle = \frac{1}{\sqrt{2}}\bigl(|j{=}3/2,\mu\rangle + (-1)^b
   |j{=}1/2,\mu\rangle\bigr) \, ,
\end{equation}
where $b=1,2$ and $\mu=1,\ldots,4$.  Alice can encode 3 c-bits into
these eight states, which are completely distinguishable by Bob.  It
is easily shown using Eq.~(\ref{eq:E3Decomp}) that the decohering
superoperator $\mathcal{E}_3$ maps all of these states to the
completely mixed state on the total Hilbert space; thus, these states
are completely indistinguishable by Eve.  This scheme is optimally
efficient for private classical communication because, by Holevo's
theorem, three c-bits is the maximum amount of classical communication
that can be achieved with three transmitted qubits.

So we see that the optimal efficiency for private classical
communication (three c-bits) is greater than that for private
quantum communication (two qubits) if we directly compare c-bits
to qubits. This result generalizes in the case of $N$ transmitted
qubits. Note, however, that the ratio of private capacity to
public capacity decreases with increasing $N$.

The examples presented in this section illustrate the central concepts
of this paper.  We now turn to the general case.

\section{Private quantum communication}
\label{sec:Quantum}

\subsection{General schemes for private quantum communication}

We begin the general discussion by defining private quantum
communication schemes (using public quantum channels and without
classical ``broadcast'' channels) as in~\cite{Amb00}, and deriving
some general results for such schemes.

Any time Alice and Bob have some private shared correlation, that is,
one to which Eve does not have access, Eve's description of the
systems transmitted along the channel is related to Alice's
description by a decohering superoperator, denoted by $\mathcal{E}$.

\textbf{Definition:} \emph{A private quantum communication scheme for
  $\mathcal{E}$.}  Such a scheme consists of an \emph{encoding}
$\mathcal{C}$, mapping message states in a logical Hilbert space
$\mathbb{H}_L$ to encoded states on the Hilbert space $\mathbb{H}$ of
the transmitted system, such that (i) the map $\mathcal{C}$ is
invertible by Bob (who possesses the private shared correlations),
allowing him to decode and recover states on $\mathbb{H}_L$ with
perfect fidelity, and (ii) the encoding satisfies
\begin{equation}
  \label{eq:EncodedStates}
  \mathcal{E}[\mathcal{C}(\varrho_L)] = \rho_0 \, , \quad \forall\
  \varrho_L\ \text{on}\ \mathbb{H}_L \, ,
\end{equation}
where $\rho_0$ is some fixed state on $\mathbb{H}$.  This latter
property ensures that all encoded states are completely
indistinguishable from Eve's perspective, so that she cannot acquire
any information about $\varrho_L$ through measurements on
$\mathcal{E}[\mathcal{C}(\varrho_L)]$.

This definition is equivalent to a ``private quantum channel'' defined
in~\cite{Amb00}.  We define an \emph{optimally efficient} private
quantum communication scheme as one for which $\mathbb{H}_L$ is of
maximal dimension.

The invertibility of the encoding $\mathcal{C}$ by Bob places
stringent conditions on the image of the logical Hilbert space
$\mathbb{H}_L$ in $\mathbb{H}$.  In order to ensure this
invertibility, one method of encoding is to choose $\mathcal{C}$ such
that $\mathbb{H}_L$ maps isomorphically to a subspace $\mathbb{H}'
\subset \mathbb{H}$ of equal dimension.  However, the most general
method of encoding involves using ancilla systems~\cite{Amb00}.  Let
$\mathbb{H}'' \subset \mathbb{H}$ be a subspace that possesses a
tensor product structure $\mathbb{H}'' = \mathbb{H}_A \otimes
\mathbb{H}_B$ with $\mathbb{H}_A$ isomorphic to $\mathbb{H}_L$.  The
Hilbert space $\mathbb{H}_A$ is referred to as a \emph{subsystem} of
$\mathbb{H}$.  An encoding $\mathcal{C}$ that maps any state
$\varrho_L$ on $\mathbb{H}_L$ to the state $\varrho_L \otimes
\sigma_0$ on $\mathbb{H}_A \otimes \mathbb{H}_B$ for some fixed
ancillary state $\sigma_0$ on $\mathbb{H}_B$ is the most general
encoding that is invertible.  In this case, we say that $\mathbb{H}_L$
is encoded by $\mathcal{C}$ into the subsystem $\mathbb{H}_A$.

In order for encoded states in a subsystem to be completely
indistinguishable by Eve, the superoperator $\mathcal{E}$ must map
them all to the same density matrix $\rho_0$ on $\mathbb{H}$.  We give
a name to such subsystems.

\textbf{Definition:} \emph{Completely private subystems.}  For all
$\varrho_L$ on $\mathbb{H}_A$, and for a fixed $\sigma_0$ on
$\mathbb{H}_B$, if
\begin{equation}
  \label{eq:EncodedStatesInSubsystem}
  \mathcal{E}(\varrho_L \otimes \sigma_0) = \rho_0 \, ,
\end{equation}
where $\rho_0$ is independent of $\varrho_L$, then the subsystem
$\mathbb{H}_A$ is said to be \emph{completely private} with respect to
$\mathcal{E}$.

Every completely private subsystem with respect to a superoperator
$\mathcal{E}$ allows for the definition of a private quantum
communication scheme.  The scheme simply encodes a logical Hilbert
space isomorphically into this completely private subsystem.

\subsection{Decoherence-full subsystems}
\label{subsec:DFull}

In the following, we highlight a particular class of completely
private subsystems, namely, those for which every state defined on the
subsystem is mapped by $\mathcal{E}$ to the completely mixed state on
the subsystem.  In contrast to the decoherence-free (D-free) or
noiseless subsystems~\cite{Kni00,Zan01} employed in quantum computing,
the effect of the decoherence on these subsystems is maximal, and so
we dub these \emph{decoherence-full} (D-full) subsystems.

\textbf{Definition:} \emph{Decoherence-full subspaces/subsystems.}
Consider a superoperator $\mathcal{E}$ that acts on density operators
on a Hilbert space $\mathbb{H}$.  A \emph{decoherence-full (D-full)
  subspace} is a subspace $\mathbb{H}' \subset \mathbb{H}$ such that
the superoperator $\mathcal{E}$ maps every density operator on
$\mathbb{H}'$ to the completely mixed density operator on
$\mathbb{H}'$.  Consider a subspace $\mathbb{H}'' \subset
\mathbb{H}$ that possesses a tensor product structure
$\mathbb{H}'' = \mathbb{H}_A \otimes \mathbb{H}_B$ such that
\begin{equation}
  \label{eq:EonPair}
  \mathcal{E}(\rho_A \otimes \rho_B) = \tfrac{1}{d_A}I_A \otimes \rho_B' \, ,
\end{equation}
where $\tfrac{1}{d_A} I_A$ is the completely mixed state on
$\mathbb{H}_A$ and $\rho_B'$ is independent of $\rho_A$. We define
such a $\mathbb{H}_A$ to be a \emph{decoherence-full (D-full)
subsystem}. If, in addition, $\rho'_B=\rho_B$ for all $\rho_B$, so
that $\mathbb{H}_B$ is decoherence-free, that is, if
\begin{equation}
  \label{eq:EonPair2}
  \mathcal{E}(\rho_A \otimes \rho_B) = \tfrac{1}{d_A}I_A \otimes \rho_B \, ,
\end{equation}
for all $\rho_A \otimes \rho_B$, then we define the product
$\mathbb{H}_A \otimes \mathbb{H}_B$ to be a \emph{D-full/D-free
  subsystem pair}.  Restricted to a D-full/D-free subsystem pair, the
superoperator $\mathcal{E}$ has the decomposition $\mathcal{E}_{AB} =
\mathcal{D}_A \otimes \mathcal{I}_B$ with respect to this TPS, where
$\mathcal{D}_{A}$ is the completely depolarizing superoperator on
$\mathbb{H}_{A}$, and $\mathcal{I}_{B}$ acts trivially on
$\mathbb{H}_{B}$.

Note that a D-full subspace is a special case of a D-full subsystem
for which $\mathbb{H}_B$ is one-dimensional.

In the following, we will show that D-full subsystems define optimally
efficient schemes for private quantum communication for the class of
superoperators describing Eve's ignorance of an SRF.

\subsection{Group-averaging superoperators}

The results so far in this section have not made any assumptions about
the sort of private shared correlation that Alice and Bob are using to
encode their information.  We now focus on the case of a private SRF.
This restriction will allow for a simple decomposition of the total
Hilbert space into D-full/D-free subsystem pairs.

Note first that every reference frame is associated with a symmetry
group.  For instance, a Cartesian frame is associated with the group
of rotations SU(2), a clock (phase reference) is associated with U(1),
and a reference ordering (which we shall consider in section
\ref{sec:quantumschemeprivaterefordering}) is associated with the
symmetric group $S_N$~\footnote{However, note that a \emph{partial}
  reference frame is associated with a factor space of a group; e.g.,
  a reference direction is associated with the factor space
  SU(2)/U(1), where U(1) is the symmetry group of the direction under
  rotations.}.  If Eve does not share Alice and Bob's RF, then she is
ignorant of which element of the group describes the relation between
her local RF and that of Alice and Bob.  The unital superoperator
$\mathcal{E}$ describing Eve's ignorance is therefore an average over
the collective representation $T$ of a group $G$ acting on
$\mathbb{H}$. If $G$ is a Lie group, then $\mathcal{E}$ acts on states
$\rho$ on $\mathbb{H}$ as
\begin{equation}
  \label{eq:SuperopAverageLieGroup}
  \mathcal{E}(\rho) = \int_G {\rm d}v(g) T(g) \rho T^\dag(g) \, ,
\end{equation}
where ${\rm d}v$ is the group-invariant measure on $G$.  For finite
groups, the superoperator acts as
\begin{equation}
  \label{eq:SuperopAverageFiniteGroup}
  \mathcal{E}(\rho) = \frac{1}{{\rm dim}\ G} \sum_i T(g_i) \rho
  T^\dag(g_i) \, ,
\end{equation}
where ${\rm dim}\ G$ is the dimension of $G$. In the following, we use
the notation of Lie groups; all results are equally applicable to
finite groups.

If $T$ acts irreducibly on $\mathbb{H}$, then $\mathcal{E}$ is
completely depolarizing (by Schur's lemma).  However, if $T$ is
reducible, then we can use the irreducible representations (irreps)
$T_j$ of $G$ to construct projection operators
\begin{equation}
  \label{eq:ProjOperators}
  \Pi_j \propto \int_G {\rm d}v(g) T_j(g^{-1}) T(g) \, ,
\end{equation}
up to a constant of proportionality.  These projection operators
decompose the Hilbert space $\mathbb{H}$ into a direct sum as
\begin{equation}
  \label{eq:GeneralDirectSum}
  \mathbb{H} = \bigoplus_j \mathbb{H}_j \, .
\end{equation}
In general, each irrep occurs multiple times; we can factor each
subspace $\mathbb{H}_j$ into a tensor product of subspaces
$\mathbb{H}_{jA} \otimes \mathbb{H}_{jB}$ as follows.  Each subsystem
$\mathbb{H}_{jA}$ is the carrier space for the irreducible
representation $T_j$ of $G$, and each corresponding subsystem
$\mathbb{H}_{jB}$ carries the trivial representation of $G$ and has
dimension equal to the multiplicity of $T_j$.  (See~\cite{Ful91}.)
The total Hilbert space decomposes as
\begin{equation}
  \label{eq:GeneralDirectSumProd}
  \mathbb{H} = \bigoplus_j \mathbb{H}_{jA} \otimes \mathbb{H}_{jB} \, .
\end{equation}
Each subsystem $\mathbb{H}_{jA}$ is D-full, and each subsystem
$\mathbb{H}_{jB}$ is D-free.  Thus, each $\mathbb{H}_{jA} \otimes
\mathbb{H}_{jB}$ form a D-full/D-free subsystem pair.  The action of the
superoperator $\mathcal{E}$ can be expressed in terms of
this decomposition as
\begin{equation}
  \label{eq:ActionOfEOnArbitrary}
  \mathcal{E}(\rho) = \sum_j (\mathcal{D}_{jA} \otimes
  \mathcal{I}_{jB}) (\Pi_j \rho \Pi_j) \, ,
\end{equation}
where $\mathcal{D}_{jA}$ is the completely depolarizing superoperator
on each $\mathbb{H}_{jA}$, and $\mathcal{I}_{jB}$ acts trivially on
each $\mathbb{H}_{jB}$.

It should be noted that the $\mathbb{H}_{jA}$ are the \emph{only}
D-full subsystems. This claim follows from the fact that if a
subsystem is D-full then the representation $T$ of $G$ must act
irreducibly when restricted to it, and the fact that the
$\mathbb{H}_{jA}$ are the only subsystems on which the representation
$T$ of $G$ acts irreducibly.  The inference from a subsystem being
D-full to having $T$ act irreducibly upon it is perhaps not obvious,
so we give a short proof by contradiction.  Suppose $\mathbb{H}_{A}$
is a D-full subsystem on which $T$ acts reducibly. It then follows
that there exists an invariant subspace $\mathbb{H}_A' \subset
\mathbb{H}_A$, meaning that for any $g \in G$, $T(g)$ maps
$\mathbb{H}_A'$ onto itself.  Thus, the action of $\mathcal{E}$ of
Eq.~(\ref{eq:SuperopAverageLieGroup}) must take a state in
$\mathbb{H}_A'$ to a state with support entirely on $\mathbb{H}_A'$,
which cannot be the completely mixed state on $\mathbb{H}_A$.  It
follows that $\mathbb{H}_A$ is not a D-full subsystem, which
contradicts our initial assumption.

\subsection{Optimally efficient private quantum communication schemes}

We can now prove our central result for private quantum communication
schemes:

\textbf{Theorem 1:} An optimally efficient private quantum
communication scheme for a group-averaging decohering superoperator
$\mathcal{E}$ is given by encoding into the largest D-full subsystem
for $\mathcal{E}$.

\textbf{Proof:} It is clear that every private quantum communication
scheme encodes into a completely private subsystem. It suffices
therefore to show that the dimension of any completely private
subsystem for a group-averaging decohering superoperator $\mathcal{E}$
is less than or equal to the dimension of the largest D-full subsystem
for $\mathcal{E}$.

Let $\mathbb{H}_E$ be a completely private subsystem for a
group-averaging decohering superoperator $\mathcal{E}$ of the form
given in Eq.~(\ref{eq:SuperopAverageLieGroup}), and let
$\mathbb{H}'_E$ be the complementary subsystem such that $\mathbb{H}_E
\otimes_E \mathbb{H}'_E \subset \mathbb{H}$ (where $\otimes_E$ denotes
the tensor product structure with respect to these subsystems).

The condition for $\mathbb{H}_E$ to be completely private is
\begin{equation}
  \label{eq:RepeatCompletelyPrivate}
  \mathcal{E}(|\psi_E\rangle\langle\psi_E| \otimes_E \sigma_0) =
  \rho_0 \, , \quad \forall\ |\psi_E\rangle \in \mathbb{H}_E \,,
\end{equation}
for some fixed state $\sigma_0$ on $\mathbb{H}'_E$, where $\rho_0$ is
a density operator on $\mathbb{H}$ that is independent of
$|\psi_E\rangle$.  Because $\sigma_0$ is arbitrary, we can choose it
to be a pure state $\sigma_0 = |\phi_0\rangle\langle\phi_0|$ for
$|\phi_0\rangle \in \mathbb{H}'_E$, which simplifies our proof.

Using the expression~(\ref{eq:ActionOfEOnArbitrary}) for the action of
$\mathcal{E}$ and projecting both sides of
condition~(\ref{eq:RepeatCompletelyPrivate}) onto an irrep $j$ gives
\begin{equation}
  \label{eq:EOnSingleIrrep}
  (\mathcal{D}_{jA} \otimes \mathcal{I}_{jB})
  (|\psi_{jE}\rangle\langle\psi_{jE}|) = \rho_{0j} \, ,
\end{equation}
where we have defined $|\psi_{jE}\rangle \equiv \Pi_j (|\psi_E\rangle
\otimes_E |\phi_0\rangle) \in \mathbb{H}_j$ and $\rho_{0j} \equiv
\Pi_j\rho_0\Pi_j$.  Consider an irrep $j$ for which $\rho_{0j}
\neq 0$.  (At least one such $j$ must exist, as the irreps span the
Hilbert space.)  Taking the partial trace over the D-full subsystem
$\mathbb{H}_{jA}$ (denoted ${\rm Tr}_{jA}$) and using the cyclic
property of trace to eliminate $\mathcal{D}_{jA}$ gives
\begin{equation}
  \label{eq:PartialTraceOnSingleIrrep}
  {\rm Tr}_{jA}
  (|\psi_{jE}\rangle\langle\psi_{jE}|) = {\rm
  Tr}_{jA}(\rho_{0j})\,, \quad \forall\ |\psi_E\rangle \in
  \mathbb{H}_E \, .
\end{equation}
Let $|\psi_E\rangle$ and $|\chi_E\rangle$ be two orthogonal states in
$\mathbb{H}_E$.  Because $\mathbb{H}_E$ is a linear space,
$(|\psi_E\rangle + |\chi_E\rangle)/\sqrt{2} \in \mathbb{H}_E$; thus,
\begin{align}
  \label{eq:PlusSuperposition}
  {\rm Tr}_{jA}(\rho_{0j})
  &= {\rm Tr}_{jA}(|\psi_{jE}\rangle\langle\psi_{jE}|) \,, \nonumber \\
  {\rm Tr}_{jA}(\rho_{0j})
  &= {\rm Tr}_{jA}(|\chi_{jE}\rangle\langle\chi_{jE}|) \,, \\
  {\rm Tr}_{jA}(\rho_{0j})
  &= \tfrac{1}{2}{\rm Tr}_{jA}\bigl[(|\psi_{jE}\rangle +
  |\chi_{jE}\rangle)(\langle\psi_{jE}| +
  \langle\chi_{jE}|)\bigr] \, . \nonumber
\end{align}
These equations lead to the identity
\begin{equation}
  \label{eq:PlusIdentity}
  {\rm Tr}_{jA}(|\psi_{jE}\rangle\langle\chi_{jE}|) +
  {\rm Tr}_{jA}(|\chi_{jE}\rangle\langle\psi_{jE}|)=0\, .
\end{equation}
Repeating this argument for $(|\psi_E\rangle + i
|\chi_E\rangle)/\sqrt{2} \in \mathbb{H}_E$ gives
\begin{equation}
  \label{eq:MinusIdentity}
  {\rm Tr}_{jA}(|\psi_{jE}\rangle\langle\chi_{jE}|) -
  {\rm Tr}_{jA}(|\chi_{jE}\rangle\langle\psi_{jE}|)=0 \, .
\end{equation}
Combining these equations, we obtain
\begin{equation}
  \label{eq:PartialTraceOnSingleIrrep2}
  {\rm Tr}_{jA}
  (|\psi_{jE}\rangle\langle\chi_{jE}|) = 0\, \quad \text{if}\
  \langle\psi_E|\chi_E\rangle = 0 \, .
\end{equation}
Let $|\zeta_{jB}\rangle$ be any state in $\mathbb{H}_{jB}$ such that
$\langle\zeta_{jB}| \rho_{0j} |\zeta_{jB}\rangle \neq 0$ (guaranteed
to exist if $\rho_{0j} \neq 0$).  We define the relative state of
$|\zeta_{jB}\rangle$ with respect to $|\psi_{jE}\rangle$, denoted
$|\psi_{jE,A}\rangle$, by
\begin{equation}
  \label{eq:ReducedState}
  |\psi_{jE,A}\rangle \equiv \langle\zeta_{jB}| \psi_{jE}\rangle
  \,.
\end{equation}
All such relative states are nonzero because
\begin{align}
  \label{eq:NonZeroReducedStates}
  \langle\psi_{jE,A}|\psi_{jE,A}\rangle &= {\rm Tr}_{jA}
  \bigl( \langle\zeta_{jB} |
  \psi_{jE}\rangle \langle\psi_{jE}|\zeta_{jB}\rangle \bigr)
  \nonumber \\
  &= \langle\zeta_{jB}| {\rm
  Tr}_{jA}(|\psi_{jE}\rangle\langle\psi_{jE}|)|\zeta_{jB}\rangle
  \nonumber \\ &=
  \langle\zeta_{jB}| {\rm
  Tr}_{jA}(\rho_{0j}) |\zeta_{jB}\rangle \nonumber
  \\
  &\neq 0 \, ,
\end{align}
where the third equality uses
Eq.~(\ref{eq:PartialTraceOnSingleIrrep}).  The relative states of
$|\zeta_{jB}\rangle$ with respect to a pair of orthogonal states,
$|\psi_E\rangle$ and $|\chi_E\rangle$, in $\mathbb{H}_E$ satisfy
\begin{align}
  \label{eq:OrthogonalReducedStates}
  \langle\chi_{jE,A}|\psi_{jE,A}\rangle
  &= {\rm Tr}_{jA} \bigl( \langle\zeta_{jB} |
  \chi_{jE}\rangle \langle\psi_{jE}|\zeta_{jB}\rangle \bigr)
  \nonumber \\
  &= \langle\zeta_{jB}| {\rm
  Tr}_{jA}(|\psi_{jE}\rangle\langle\chi_{jE}|)|\zeta_{jB}\rangle
  \nonumber \\ &= 0 \, ,
\end{align}
where the final step follows from
Eq.~(\ref{eq:PartialTraceOnSingleIrrep2}).  Thus, for any two
orthogonal states $|\psi\rangle_E$ and $|\chi\rangle_E$ in
$\mathbb{H}_E$, there exists a pair of non-zero orthogonal states in
$\mathbb{H}_{jA}$. The number of orthogonal states in
$\mathbb{H}_{jA}$ is upper bounded by its dimension.  Thus, the
dimension of any completely private subsystem $\mathbb{H}_E$ cannot be
greater than the dimension of the D-full subsystem $\mathbb{H}_{jA}$
for any $j$ for which $\rho_{0j} \neq 0$.

It follows that the dimension of a completely private subsystem cannot
be greater than the dimension of the \emph{largest} D-full subsystem.
Thus, an optimally efficient encoding in achieved by using the largest
D-full subsystem.  \hfill$\Box$\medskip

\subsection{Optimally efficient quantum communication scheme for a
  private shared Cartesian frame}

We now use the group theoretical structure of the superoperator
$\mathcal{E}_N$ to determine the optimally efficient quantum
communication scheme for a private shared Cartesian frame and
transmission of $N$ spin-1/2 particles. The Hilbert space
$(\mathbb{C}^2)^{\otimes N}$ of these $N$ qubits carries a collective
tensor representation $R^{\otimes N}$ of SU(2), by which a rotation
$\Omega \in$ SU(2) acts identically on each of the $N$ qubits.  This
Hilbert space also carries a representation $P_N$ of the symmetric
group $S_N$, which is the group of permutations of the $N$ qubits. The
action of these two groups commute, and Schur-Weyl
duality~\cite{Ful91} states that the Hilbert space
$(\mathbb{C}^2)^{\otimes N}$ carries a multiplicity-free direct sum of
SU(2)$\times S_N$ irreps, each of which can be labelled by the SU(2)
total angular momentum quantum number $j$.  For simplicity, we
restrict $N$ to be an even integer for the remainder of this paper.
Then,
\begin{equation}
  \label{eq:DirectSum}
  (\mathbb{C}^2)^{\otimes N} = \bigoplus_{j=0}^{N/2} \mathbb{H}_j \,,
\end{equation}
where $\mathbb{H}_j$ is the eigenspace of total angular momentum with
eigenvalue $j$, and the group SU(2)$\times S_N$ acts irreducibly on
each eigenspace.

Because the groups SU(2) and $S_N$ commute, the Hilbert space can be
further decomposed.  Each subspace $\mathbb{H}_j$ in the direct sum
can be factored into a tensor product $\mathbb{H}_j = \mathbb{H}_{jR}
\otimes \mathbb{H}_{jP}$, such that SU(2) acts irreducibly on
$\mathbb{H}_{jR}$ and trivially on $\mathbb{H}_{jP}$, and $S_N$ acts
irreducibly on $\mathbb{H}_{jP}$ and trivially on $\mathbb{H}_{jR}$.
Thus,
\begin{equation}
  \label{eq:DirectSumOfProducts}
  (\mathbb{C}^2)^{\otimes N} = \bigoplus_{j=0}^{N/2} \mathbb{H}_{jR}
  \otimes \mathbb{H}_{jP} \, .
\end{equation}
The dimension of $\mathbb{H}_{jR}$ is
\begin{equation}
  \label{eq:MultiplicityR}
  d_{jR}=2j+1\, ,
\end{equation}
and that of $\mathbb{H}_{jP}$ is~\cite{BRS03a}
\begin{equation}
  \label{eq:Multiplicity}
  d_{jP}=\binom{N}{N/2-j}\frac{2j+1}{N/2+j+1} \, .
\end{equation}

If Alice prepares $N$ qubits in a state $\rho$ and sends them to Bob,
an eavesdropper Eve who is uncorrelated with the private SRF will
describe the state as mixed over all rotations $\Omega \in$ SU(2).
Thus, the superoperator $\mathcal{E}_N$ acting on a general density
operator $\rho$ of $N$ qubits that describes the lack of knowledge of
this private SRF is given by~\cite{BRS03a}
\begin{equation}
  \label{eq:NQubitDecoheringChannel}
  \mathcal{E}_N(\rho) = \int {\rm d}\Omega \, R(\Omega)^{\otimes N}
  \rho R^{\dag}(\Omega)^{\otimes N} \, .
\end{equation}
The effect of this superoperator is best seen through use of the
decomposition (\ref{eq:DirectSumOfProducts}) of the Hilbert space.
The subsystems $\mathbb{H}_{jP}$ are D-free or noiseless
subsystems~\cite{Kni00} under the action of this superoperator; states
encoded into these subsystems are completely protected from this
decoherence.  In contrast, $\mathcal{E}_N$ is completely depolarizing
on each $\mathbb{H}_{jR}$ subsystem, and thus the $\mathbb{H}_{jR}$
are D-full subsystems.  For each $j$, the subsystems $\mathbb{H}_{jR}
\otimes \mathbb{H}_{jP}$ form a D-full/D-free subsystem pair.

The largest D-full subsystem occurs for $j_{\rm max} = N/2$ and has
dimension $2j_{\rm max}+1 =N+1$.  This D-full subsystem defines the
optimally efficient private quantum communication scheme (by Theorem
1).  Thus, given a private Cartesian frame and the transmission of $N$
qubits, Alice and Bob can privately communicate $\log (N+1)$ qubits,
or $\log (N)$ qubits asymptotically.

\subsection{Optimally efficient quantum communication scheme for a
  private shared reference ordering}
\label{sec:quantumschemeprivaterefordering}

Note the duality of the rotation group and the symmetric group in
the system described above.  One may ask why we consider a
reference frame for the first group and not the second.  In fact,
we have implicitly assumed a reference frame for the permutation
group in the form of a \emph{shared reference ordering}.  The
simplest way in which two parties can possess a shared reference
ordering is if they agree on some labelling of the qubits, for
instance, using their temporal order, and if the quantum channel
preserves this labelling.  The shared reference ordering that has
been assumed up until now has been taken to be public (i.e., Eve
shares it as well); however, one can also consider it to be
private. Here, we consider the dual problem to the one of the
previous section: a public Cartesian frame and a private reference
ordering.

Note that sharing a private reference ordering is not equivalent
to sharing a secret key.  This inequivalence may seem surprising,
because the most obvious way in which Alice and Bob may share a
private reference ordering is for them to agree on a secret
permutation of $N$ elements (Alice applies the permutation to the
qubits prior to transmission and Bob applies it to the qubits
after receiving them). As there are $N!$ elements in $S_N$, this
secret permutation is equivalent to sharing $\log (N!)$ bits of
secret key.  Nonetheless, in general when Alice and Bob share a
private reference ordering they need not share any secret key. For
instance, suppose the channel that connects Alice and Bob
implements some fixed permutation $p_{C}$ of the qubits, and that
this permutation is unknown to both Alice and Bob.  The shared
reference ordering is provided to Alice and Bob in the form of two
devices, one for each party.  Alice's device applies some
permutation $p_A$ to her qubits prior to transmission, and Bob's
device applies some permutation $p_B$ upon receiving them.  The
devices are designed such that $p_B = (p_C p_A)^{-1}$, and thus
Bob recovers the quantum state of the qubits prepared by Alice.
Assuming that $p_C$ is equally likely to be any element of $S_N$,
Alice has no knowledge of $p_B$ and Bob has no knowledge of $p_A$.
Therefore, they do not share a secret key.  Note further that
although Eve may have knowledge of $p_C$ (which she may acquire,
for instance, by examining the channel), she has no knowledge of
$p_A$, and assuming that $p_A$ is chosen uniformly among elements
of $S_N$, Eve's description of the qubits is related to
Alice's description by the superoperator
\begin{equation}
  \label{eq:MixAllPermutations}
  \mathcal{P}_N[\rho] = \frac{1}{N!} \sum_{p \in S_N} P(p) \rho
  P^\dag(p) \, ,
\end{equation}
where $P(p)$ is the unitary operator corresponding to the permutation
$p$ of the qubits.

When the $P(p)$ are decomposed into irreps, $\mathcal{P}_N$ induces
the decomposition of $\mathbb{H}$ specified in
Eq.~(\ref{eq:DirectSumOfProducts}), which is the same decomposition
that was induced by $\mathcal{E}_N$.  However, there is a difference:
with respect to the superoperator $\mathcal{P}_N$, the subsystems
$\mathbb{H}_{jP}$ are D-full (because $S_N$ acts irreducibly on these
subsystems) and the subsystems $\mathbb{H}_{jR}$ are D-free (because
$S_N$ acts trivially on these).

For large $N$, the largest $\mathbb{H}_{jP}$ occurs for $j = j_{\rm
  max}$, the integer nearest to $\sqrt{N}/2$, and has dimension
$d_{jP} = O(2^N/N)$ (meaning that $d_{jP} < c 2^N/N$ for some
constant $c$ and for all values of $N$) as can be deduced from
Eq.~(\ref{eq:Multiplicity}).  This D-full subsystem defines the
optimally efficient private quantum communication scheme.  It
allows for private communication of $N - \log_2 N$ logical qubits
asymptotically given $N$ transmitted qubits.

\subsection{Optimally efficient scheme for a private shared Cartesian
  frame and reference ordering}

Another interesting case is the one where Alice and Bob possess both a
private Cartesian frame as well as a private reference ordering. For
transmission of $N$ qubits in this situation, Eve's lack of knowledge
about either reference is characterized by the superoperator
$\mathcal{E}_N \circ \mathcal{P}_N$.  Interestingly, this
superoperator is not completely depolarizing on the entire Hilbert
space.  Even without sharing either reference, Eve can still measure
the total $\hat{J}^2$ operator to acquire information about the
preparation.  However, the subspaces $\mathbb{H}_{jR} \otimes
\mathbb{H}_{jP}$ for each $j$ are D-full under the action of this
superoperator. Thus, Alice and Bob can perform private quantum
communication by encoding into one of these spaces.  The largest
D-full subspace occurs for $j = j_{\rm max}$, the integer nearest to
$\sqrt{N}/2$, and has dimension $d_{jR}d_{jP} = O(2^N/\sqrt{N})$.
Asymptotically, this allows for $N- \frac{1}{2}\log_2 N$ private
logical qubits to be encoded in $N$ transmitted qubits.

\subsection{The duality between cryptography and communication}

We have been concerned with determining how much quantum
information, prepared relative to some RF, can be completely
hidden from someone who does not share this RF.  If this person is
an eavesdropper, then this concealment can be very useful for
cryptography, as we have shown.  However, it can occur that
someone with whom one \emph{wants} to communicate does not share
the RF, for whatever reason.  In this case, one is interested in
the opposite problem, namely, how much quantum information can be
made completely \emph{accessible} to someone who does not share
the RF.  This amount is determined by the largest D-free
subsystem, as was shown in~\cite{BRS03a}.  The following dichotomy
arises: information encoded in a D-full subsystem is hidden from
someone lacking the RF, while information encoded in a D-free
subsystem is still accessible to someone lacking the RF.  The
implications for the case we are considering can be summarized as
follows.  From the perspective of someone who lacks the SU(2) SRF,
the $\mathbb{H}_{jR}$ are D-full and the $\mathbb{H}_{jP}$ are
D-free; from the perspective of someone who lacks the $S_N$ SRF,
it is the $\mathbb{H}_{jP}$ that are D-full and the
$\mathbb{H}_{jR}$ that are D-free.  Thus, the number of logical
qubits that can be transmitted privately given a private SU(2) SRF
is equal to the number of logical qubits that can be communicated
to a receiver that lacks the $S_N$ SRF, and similarly with SU(2)
and $S_N$ reversed.  What is bad for private quantum communication
using a private SRF is good for quantum communication in the
absence of an SRF.

\section{Private classical communication}
\label{sec:Classical}

We now consider the private communication of classical information
through a quantum channel using the resource of a private SRF.  We
provide upper bounds on the efficiency of such schemes (maximum number
of private messages that can be sent), and present schemes for private
SU(2) and/or $S_N$ SRFs that asymptotically saturate these bounds.  As
it turns out, the optimally efficient schemes for private classical
communication are more efficient than the optimally efficient private
quantum schemes (comparing private c-bits directly with private
qubits).

\textbf{Definition:} \emph{A private classical communication scheme
  for a decohering superoperator $\mathcal{E}$.}  Such a scheme
consists of a set $\{ \rho_i \}$ of density operators on $\mathbb{H}$
prepared by Alice that are (i) orthogonal, so that Bob can distinguish
these classical messages with certainty, and (ii) satisfy
\begin{equation}
  \label{eq:ClassicalEncodedStates}
  \mathcal{E}[\rho_i] = \rho_0 \, , \quad \forall\ \rho_i \, ,
\end{equation}
where $\rho_0$ is some fixed state in $\mathbb{H}$, ensuring that Eve
cannot gain any information about these classical messages. An
optimally efficient private classical communication scheme has the
maximum number of elements in the set $\{ \rho_i \}$.

It is clear that every private \emph{quantum} communication scheme can
be turned into a private \emph{classical} communication scheme by
encoding the classical messages into an orthogonal set of quantum
states within the D-full subsystem employed by the latter.  However,
we now show that for the group-averaging superoperators, there exist
private classical communication schemes that perform much better.  As
with our three qubit example given in Section~\ref{sec:Example}, the
key to finding efficient private classical communication schemes is to
encode into states that are entangled between D-full and D-free
subsystems and span many irreps.

\subsection{An illustrative example}

Consider the following illustrative example.  Let $\mathbb{H}$ be a
Hilbert space.  Let $\mathcal{E}$ be a superoperator acting on
states of this space such that, under a decomposition of $\mathbb{H}$
as
\begin{equation}
  \label{eq:ExampleDirectSum}
  \mathbb{H} = \bigoplus_{a=1}^A \mathbb{H}_{a1} \otimes
  \mathbb{H}_{a2} \,,
\end{equation}
the subsystems $\mathbb{H}_{a1} \otimes \mathbb{H}_{a2}$ are
D-full/D-free subsystem pairs under the action of $\mathcal{E}$.  For
our example, we enforce the additional (and atypical) constraint that
\begin{equation}
  \label{eq:DimensionsTheSame}
  {\rm dim}\,\mathbb{H}_{a1} = {\rm dim}\,\mathbb{H}_{a2} = d \, ,
\end{equation}
for some integer $d$ independent of $a$.  Thus, all of the D-full
subsystems $\mathbb{H}_{a1}$ and D-free subsystems $\mathbb{H}_{a2}$
are of the same dimension, and the dimension of the total Hilbert
space $\mathbb{H}$ is $Ad^2$.

If Eve's lack of correlations is described by the superoperator
$\mathcal{E}$, then a simple private classical communication scheme
can be constructed as follows.  For a \emph{fixed} arbitrary $a$,
choose a set of $d$ orthogonal states $\{|a,k\rangle_1, k=1,\ldots,d
\}$ spanning the D-full subsystem $\mathbb{H}_{a1}$, and an arbitrary
fixed state $|a,0\rangle_2 \in \mathbb{H}_{a2}$.  Then $d$ classical
messages can be encoded into the $d$ orthogonal states $|a,k\rangle_1
\otimes |a,0\rangle$.  All of these states map to the same density
operator $\tfrac{1}{d}I_{a1} \otimes |a,0\rangle_2\langle a, 0|$ under
the action of $\mathcal{E}$.

However, a more efficient scheme can be constructed using
\emph{entangled} states in $\mathbb{H}_{a1} \otimes \mathbb{H}_{a2}$,
as follows.

Let $\{|a,k\rangle_1, k=1,\ldots,d \}$ be a basis for
$\mathbb{H}_{a1}$, and $\{|a,k'\rangle_2, k'=1,\ldots,d \}$ be a basis
for $\mathbb{H}_{a2}$.  The states
\begin{equation}
  \label{eq:EntangledStates}
  |\psi_{alm}\rangle = \frac{1}{\sqrt{d}}\sum_{k=1}^d
   \exp(2\pi {\rm i}km/d)|a,k\rangle_1 |a,k+l\rangle_2 \, ,
\end{equation}
for $l,m=1,\ldots,d$ are an orthogonal basis of $d^2$ maximally
entangled states in $\mathbb{H}_{a1} \otimes \mathbb{H}_{a2}$.  Using
the fact that the maximally entangled states $|\psi_{alm}\rangle$
possess maximally mixed reduced density operators ${\rm
  Tr}_{a1}(|\psi_{alm}\rangle\langle\psi_{alm}|) = \frac{1}{d}
I_{a_2}$, it follows that all such maximally entangled states map
under $\mathcal{E}$ to the state
\begin{equation}
  \label{eq:EntangledStatesDecohere}
  \mathcal{E}(|\psi_{alm}\rangle\langle\psi_{alm}|) =
  \tfrac{1}{d} I_{a 1} \otimes \tfrac{1}{d} I_{a 2} \, ,
\end{equation}
for all $l,m$.  Thus, one can encode $d^2$ messages into entangled
states of this form.

Finally, we present an optimally efficient scheme which performs even
better.  Again, we define the entangled states $|\psi_{alm}\rangle$
for every $a=1,\ldots,A$ as in Eq.~(\ref{eq:EntangledStates}); these
states form an orthogonal basis for the entire Hilbert space
$\mathbb{H}$.  We then construct the Fourier transform states over the
index $a$
\begin{equation}
  \label{eq:FourierTransformedEntStates}
  |\phi_{\mu lm}\rangle = \sum_{a=1}^A \exp(2\pi{\rm i}\mu a/A)
   |\psi_{alm}\rangle \, ,
\end{equation}
for $\mu = 1,\ldots,A$.  These states are also orthogonal:
\begin{equation}
  \label{eq:FourierTransStatesOrthogonal}
  \langle \phi_{\mu lm}|\phi_{\mu'l'm'}\rangle =
  \delta_{ll'}\delta_{mm'}\delta_{\mu\mu'} \, ,
\end{equation}
and each has the same and equal support on each of the subspaces
$\mathbb{H}_{a1} \otimes \mathbb{H}_{a2}$.  It is easily shown that
they all map under the action of $\mathcal{E}$ to the completely mixed
operator on $\mathbb{H}$; that is,
\begin{equation}
  \label{eq:MaximallyMixed}
  \mathcal{E}(|\phi_{\mu lm}\rangle\langle\phi_{\mu lm}|) =
  \tfrac{1}{Ad^2} I \,
  , \quad \forall\ l,m,\mu \, .
\end{equation}
Thus, these orthogonal states define a private classical communication
scheme.  We note that there are $Ad^2 = {\rm dim}\,\mathbb{H}$ such
states; therefore by Holevo's theorem this scheme is optimally
efficient.

The difficulty with generalizing this scheme to typical
group-averaging superoperators is that the induced tensor product
structure of D-full and D-free subsystems for a given irrep typically
do not have equal dimensions, and these change as we vary over irreps.
Below, we formulate and prove several theorems that allow us to place
upper bounds on the number of private classical messages, and to
construct asymptotically-optimal schemes for private classical
communication using private SU(2) and $S_N$ SRFs.

\subsection{A single D-full/D-free subsystem pair}
\label{sec:classicalschemeoneirrep}

Consider a decohering superoperator of the form $\mathcal{D}_{jA}
\otimes \mathcal{I}_{jB}$ defined on $\mathbb{H}_A \otimes
\mathbb{H}_B$.  This superoperator takes any state $\rho_{AB}$ on
$\mathbb{H}_A \otimes \mathbb{H}_B$ to $\tfrac{1}{d_A} I_A \otimes {\rm
  Tr}_A(\rho_{AB})$. We therefore have a single D-full/D-free
subsystem pair.  We now prove a lemma for the optimally efficient
private classical communication scheme in this case.

\textbf{Lemma 1:} Consider a Hilbert space $\mathbb{H}_A \otimes
\mathbb{H}_B$, where $\mathbb{H}_A$ ($\mathbb{H}_B$) has
dimensionality $d_A$ ($d_B$).  Let $\{ \rho_i \}$ be a private
classical communication scheme for the superoperator $\mathcal{D}_{jA}
\otimes \mathcal{I}_{jB}$. The maximum number of private classical
messages (i.e., the maximum cardinality of the set $\{ \rho_i \}$) is
$M = d_A \cdot {\rm min}\{d_A,d_B\}$.

\textbf{Proof:} We consider two separate cases for the dimensions of
the D-full and D-free subsystems.  Each proof gives a construction for
an optimally efficient private classical communication scheme.  Let
$\{ |k\rangle_A\}$ and $\{ |k\rangle_B\}$ be an orthonormal basis for
$\mathbb{H}_A$ and $\mathbb{H}_B$, respectively.

\textit{Case 1:} $d_A \geq d_B$.  The $d_A \cdot d_B$ orthogonal
maximally-entangled states
\begin{equation}
  \label{eq:OneIrrepMaxEntangled}
  |\psi_{lm}\rangle = \frac{1}{\sqrt{d_B}} \sum_{k=1}^{d_B}
   \exp(2\pi{\rm i}km/d_B) |k+l\rangle_A |k\rangle_B \, ,
\end{equation}
where $l = 1,\ldots,d_A$ and $m=1,\ldots,d_B$, satisfy
$\mathcal{D}_{jA} \otimes
\mathcal{I}_{jB}(|\psi_{lm}\rangle\langle\psi_{lm}|) = \frac{1}{d_A}
I_A \otimes \tfrac{1}{d_B} I_B$.  Thus, this set of states forms a
private classical communication scheme.  Because $d_A \cdot d_B$ is
the dimension of $\mathbb{H}_A \otimes \mathbb{H}_B$, there cannot
exist a larger set of orthogonal states on this space, and thus this
scheme is optimally efficient.

\textit{Case 2:} $d_A < d_B$.  The $d_A^2$ orthogonal
maximally-entangled states
\begin{equation}
  \label{eq:OneIrrepMaxEntangled2}
  |\psi_{lm}\rangle = \frac{1}{\sqrt{d_A}} \sum_{k=1}^{d_A}
   \exp(2\pi{\rm i}km/d_A) |k+l\rangle_A |k\rangle_B \, ,
\end{equation}
where $l,m = 1,\ldots,d_A$, satisfy $\mathcal{D}_{jA} \otimes
\mathcal{I}_{jB}(|\psi_{lm}\rangle\langle\psi_{lm}|) = \frac{1}{d_A}
I_A \otimes \sigma_B$, with
\begin{equation}
  \label{eq:PartialIdentity}
  \sigma_B = \frac{1}{d_A} \sum_{k=1}^{d_A} |k\rangle_B\langle k| \, .
\end{equation}
Thus, this set of states forms a private classical communication
scheme.

This set of states has cardinality less than the dimension of the
joint Hilbert space; however, as we now show, the scheme is optimally
efficient.  First, consider sets of pure states.  Every such state
must have the same reduced density operator on $\mathbb{H}_B$, which
we denote by $\sigma_B$.  For a pure state, the rank of the reduced
density operators on $A$ and $B$ must be equal, and because the former
is bounded above by $d_A$, the latter must be as well. Thus, we can
limit our consideration to the subspace $\mathbb{H}'_B \subset
\mathbb{H}_B$ spanned by the support of $\sigma_B$, whose dimension is
bounded above by $d_A$. But this is just \textit{Case 1} applied to
$\mathbb{H}_A \otimes \mathbb{H}'_B$, for which $d_A^2$ is the maximum
number of private messages.

It remains to be shown that making use of a set of \emph{mixed} states
does not allow for a better scheme. Imagine a set $\{ \rho_i \}$ of
mixed states on $\mathbb{H}_A \otimes \mathbb{H}_B$, containing $M$
elements. Each $\rho_i$ must have the same reduced density operator on
$\mathbb{H}_B$, which we denote by $\sigma_B$. We denote the rank of
$\sigma_B$ by $r$.  Expressing each $\rho_i$ as an eigendecomposition,
we have
\begin{equation}
  \label{eq:MixedEigendecomp}
  \rho_i = \sum_{l=1}^{L_i} p^i_l |\psi^{(i)}_l\rangle_{AB}\langle
  \psi^{(i)}_l| \, ,
\end{equation}
where $\{|\psi^{(i)}_l\rangle_{AB}|l=1,\dots,L_i\}$ are $L_i$ pure
states on $\mathbb{H}_A \otimes \mathbb{H}_B$.  Each of these pure
states has a reduced density matrix $\sigma^{(i)}_{lB} = {\rm
  Tr}_A(|\psi^{(i)}_l\rangle_{AB}\langle\psi^{(i)}_l|)$ with rank
$r_l^{(i)} \le d_A$. For each $i$, a convex sum over $l$ of the
$\sigma^{(i)}_{lB}$ must yield $\sigma_B$.  It follows that
$\sum_{l=1}^{L_i} r_l^{(i)} \ge r$, which implies that for all $i$,
$L_i d_A \ge r$. However, if all $\rho_i$ are orthogonal, they must
possess orthogonal supports, and these will therefore span a space of
dimension $\sum_{i=1}^M L_i$.  This space is contained in
$\mathbb{H}_A \otimes \mathbb{H}^{(r)}_{B}$ (where
$\mathbb{H}^{(r)}_{B} \subset \mathbb{H}_B$ is the $r$-dimensional
space spanned by the support of $\sigma_B$) and thus $\sum_{i=1}^M L_i
\le d_A r$.  Combining these inequalities yields $M \leq d_A^2$.

The set of states in Eq.~(\ref{eq:OneIrrepMaxEntangled2}) consists of
$M=d_A^2$ elements, therefore the scheme involving these states is
optimally efficient.  \hfill$\Box$\medskip

\subsection{A general group-averaging superoperator}

In the previous section we considered communication schemes using
states that are confined to a single D-full/D-free subsystem pair.
The most general scheme, however, makes use of states that span
many such pairs. We must therefore consider the more general
group-averaging superoperator $\mathcal{E}$ of
Eq.~(\ref{eq:ActionOfEOnArbitrary}).  Let $\{ \rho_i \}$ be a
private classical communication scheme for this superoperator,
satisfying $\mathcal{E}(\rho_i) = \rho_0$ for all $\rho_i$. We now
prove a lemma which bounds the cardinality of $\{ \rho_i \}$.

\textbf{Lemma 2:} An upper bound on the number of states on
$\mathbb{H}$ in a private classical communication scheme for
$\mathcal{E}$ is $M = \sum_j M_j$, where $M_j$ is the maximum number
of states on $\mathbb{H}_j$ in a private classical communication
scheme for $\mathcal{D}_{jA} \otimes \mathcal{I}_{jB}$.

\textbf{Proof:} By assumption,
\begin{equation}
  \mathcal{E}(\rho_i) = \rho_0 \, ,
\end{equation}
for all $i$. Projecting both sides of this equation onto an irrep $j$,
we obtain
\begin{equation}
  \label{eq:PrivateInOneIrrep}
  (\mathcal{D}_{jA} \otimes \mathcal{I}_{jB})(\Pi_j \rho_i \Pi_j) =
  \Pi_j \rho_0 \Pi_j \, ,
\end{equation}
for all $i$. By Lemma 1, there are at most $M_j$ orthogonal states
that are mapped by $\mathcal{D}_{jA} \otimes \mathcal{I}_{jB}$ to the
same density operator.  Therefore the supports of $\{ \Pi_j \rho_i
\Pi_j, i=1,2,\ldots \}$ must lie in a subspace of $\mathbb{H}_j$ with
dimension not greater than $M_j$.

The set of states $\{\rho_i\}$ must therefore have support on a
subspace with dimension $M = \sum_j M_j$.  The cardinality of the set
of orthogonal states $\{ \rho_i \}$ forming a private communication
scheme is therefore upper-bounded by $M = \sum_j M_j$.
\hfill$\Box$\medskip

Thus, we have the following theorem:

\textbf{Theorem 2:} In a private classical communication scheme for a
group-averaging superoperator $\mathcal{E}$, the number $M$ of private
classical messages satisfies
\begin{equation}
  \label{eq:UpperBoundOnM}
  M \le \sum_j d_{jA} \cdot {\rm min}\{d_{jA},d_{jB}\} \, ,
\end{equation}
where the $d_{jA}$ ($d_{jB}$) are the dimensions of the D-full
(D-free) subsystems defined by $\mathcal{E}$.

The proof is immediate from the preceding Lemmas.

Given that our theorem yields only an upper bound on the number of
private classical messages that can be sent, the question of exactly
how many private classical messages can be achieved remains open.  As
the example provided in Eq.~(\ref{eq:fourprivatestates}) of section
\ref{sec:twotransmittedqubits} illustrates, the optimally efficient
scheme is likely to make use of states that span irreps possessing
unequal dimensions.

\subsection{Private classical communication using a private Cartesian frame}

We now consider the specific case of a private Cartesian frame, and
present a scheme for private classical communication that is optimally
efficient in the limit of large $N$.

Consider the decomposition of the $N$-qubit Hilbert space
$(\mathbb{C}^2)^{\otimes N}$ into a direct sum of D-full/D-free
subsystem pairs as in Eq.~(\ref{eq:DirectSumOfProducts}).  First, we
note, from Eq.~(\ref{eq:MultiplicityR}) and (\ref{eq:Multiplicity}),
that for all $j$ strictly less than the maximum value $N/2$, the
D-free subsystem $\mathbb{H}_{jP}$ is always of \emph{greater or
  equal} dimension than the D-full subsystem $\mathbb{H}_{jR}$.  Thus,
we will employ irreps up to, but \emph{not} including, $j=N/2$.  Let
$j_{\rm min}<N/2$ be some fixed irrep.  We now construct orthogonal
entangled states for every irrep in the range $j_{\rm min} \leq j <
N/2$ as follows.  For convenience, we denote the dimension of the
D-full subsystem of the $j_{\rm min}$ irrep by $d$, that is,
$d\equiv2j_{\rm min}+1$.  Choose a set of orthogonal states $\{
|j,s\rangle_R, s=1,\ldots,d \}$ for $\mathbb{H}_{jR}$ and a
corresponding set of orthogonal states $\{ |j,s'\rangle_P,
s'=1,\ldots,d \}$ for $\mathbb{H}_{jP}$; note that such sets always
exist because ${\rm dim}\,\mathbb{H}_{jR} = 2j+1 \geq d$ for all $j$
in the range $j_{\rm min} \leq j < N/2$.  For each irrep in this
range, a set of $d^2$ orthogonal entangled states are then given by
\begin{equation}
  \label{eq:SU(2)MaxEntStates2}
  |\psi_{jkl}\rangle = \frac{1}{\sqrt{d}} \sum_{s=0}^d
   \exp(2\pi{\rm i}sk/d) |j,s\rangle_R |j,s+l\rangle_P \, .
\end{equation}
We wish to construct Fourier transformed states over $j$ with equal
weight in each irrep.  Thus, we define
\begin{equation}
  \label{eq:SU(2)FourierTransformedEntStates}
  |\phi_{\mu kl}\rangle = \sum_{j=j_{\rm min}}^{N/2-1} \exp(2\pi{\rm
  i}\mu j/(N/2 - j_{\rm min})) |\psi_{jkl}\rangle \, .
\end{equation}
These states are all orthogonal, and all map to the same density
matrix under the superoperator $\mathcal{E}_N$.  The range of both $i$
and $l$ is $(1,\ldots,d=2j_{\rm min}+1)$, and the range of $\mu$ is
$(1,\ldots,N/2-j_{\rm min})$; thus, there are a total of
\begin{equation}
  \label{eq:TotalOrthoFourierStates}
  M = (N/2-j_{\rm min})(2j_{\rm min}+1)^2
\end{equation}
distinct states.  To maximize this number asymptotically, we choose
$j=j_{\rm min}$, the integer nearest to $N/3$; this choice results in
$O(N^3)$ distinct states.  Thus, asymptotically, this scheme allows
for $3\log_2 N$ private classical bits to be communicated using $N$
transmitted qubits, which saturates the upper bound given by Theorem
2.

\subsection{Private classical communication using a private reference
  ordering}

As a second example of private classical communication, we consider
the case where the private SRF is a private reference ordering. As
discussed in section \ref{sec:quantumschemeprivaterefordering}, in
this case the superoperator is $\mathcal{P}_N$ (defined in
Eq.~(\ref{eq:MixAllPermutations})), and the $\mathbb{H}_{jP}$ are
D-full subsystems while the $\mathbb{H}_{jR}$ are D-free subsystems.
We consider only the limit of large $N$. In this case, the upper bound
on the number of messages is simply $2^N$, the dimensionality of the
entire Hilbert space.  This bound is saturated asymptotically by a
scheme similar to the one used in the previous section. For $j < N/2$,
we have $d_{jP}\geq d_{jR}$, so that Case 1 of the proof of Lemma 1
applies and we can define $d_{jR}d_{jP}$ entangled states within the
$j$ irrep (using Eq.~(\ref{eq:OneIrrepMaxEntangled})) which cannot be
distinguished by Eve.  For every $j$ value in a window of approximate
width $\sqrt{N}$ centered at the integer nearest $\sqrt{N}$, we
have in the asymptotic limit that $d_{jR} = O(\sqrt{N})$ and $d_{jP}=
O(2^N/N)$ (using Eqs.~(\ref{eq:MultiplicityR}) and
(\ref{eq:Multiplicity}) and Stirling's formula).  Thus, in each such
irrep, one can find $M_j = O(2^N/\sqrt{N})$ orthogonal states that
cannot be distinguished by Eve.  We can therefore Fourier transform
these states across the $\sqrt{N}$ irreps, using the construction of
Eq.~(\ref{eq:FourierTransformedEntStates}).  The end result is a set
of states that cannot be distinguished by Eve, the cardinality of
which is $M= O(2^N)$.  Thus, asymptotically, one achieves $N$ private
c-bits using this scheme.

\subsection{Private classical communication using a private Cartesian
  frame and reference ordering}

If Alice and Bob possess both a private Cartesian frame and a private
reference ordering of the transmitted qubits, then they can encode at
least as many classical messages as they could with just a private
reference ordering.  Thus, asymptotically, they can achieve $N$
private c-bits in this case as well.  One cannot achieve any more than
this, because Holevo's theorem ensures that using $N$ transmitted qubits
at most $N$ c-bits, whether private or public, can be communicated.

\section{Discussion}
\label{sec:Conclusions}

In this paper, we have demonstrated that private shared reference
frames are a resource of private correlations which can be used for
cryptography.  We have presented optimally efficient schemes for
private quantum and classical communication using an insecure quantum
channel for spin-1/2 systems and a shared Cartesian reference frame
and/or a shared reference ordering of the systems. The results are
summarized in Table~\ref{tab:Capacity}.

\begin{table}[t]
\caption{Asymptotic capacity for private quantum and classical
  communication for $N$ transmitted qubits and various private shared
  reference frames.
\label{tab:Capacity}}
\begin{center}
\footnotesize
\begin{tabular}{|c|c|c|}
\hline
\raisebox{0pt}[13pt][7pt]{Nature of the private SRF}
&\begin{minipage}{0.9in}{Private quantum}\\{capacity (qubits)}\end{minipage}&
\begin{minipage}{0.9in}{Private classical}\\{capacity (c-bits)}
\end{minipage}\\
\hline
\hline
\begin{minipage}{1.4in}{Private Cartesian frame}\\{(Private SU(2) SRF)}
\end{minipage}
&\raisebox{0pt}[13pt][7pt]{$\log_2 (N)$} &
\raisebox{0pt}[13pt][7pt]{$3\log_2 (N)$}\\
\hline
\begin{minipage}{1.4in}{Private reference ordering}\\{(Private $S_N$ SRF)}
\end{minipage}
&\raisebox{0pt}[13pt][7pt]{$N - \log_2 (N)$} &
\raisebox{0pt}[13pt][7pt]{$N$}\\
\hline
\begin{minipage}{1.4in}{Both private}\\{(Private SU(2) \& $S_N$ SRF)}
\end{minipage}
&\raisebox{0pt}[13pt][7pt]{$N - \frac{1}{2} \log_2 (N)$} &
\raisebox{0pt}[13pt][7pt]{$N$}\\
\hline
\end{tabular}
\end{center}
\end{table}

We note that our private classical schemes using a private SRF are
similar in some ways to private-key cryptography, specifically, the
Vernam cipher (one-time pad)~\cite{Ver26}.  For example, the secret
key in the Vernam cipher can be used only once to ensure perfect
security.  Similarly, for our classical schemes, only a single
plain-text (classical or quantum) can be encoded using a single
private SRF.  If the same private SRF is used to encode two
plain-texts, then the \emph{relation} that holds between the two
cipher-texts carries information about the plain-texts, and because it
is possible to learn about this relation without making use of the
SRF, Eve can obtain this information.  This fact is clear from the
example of a classical communication scheme by transmission of a
classical pencil or gyroscope, considered in the introduction.
Although Eve cannot determine the Euler angles of the pencils relative
to the shared Cartesian frame, she can measure the angular separation
of the two pencils.

It is also useful to consider the \emph{differences} between using
private shared reference frames and a secret key for private
communication.  One clear difference is that a secret key may be
subdivided into a number of smaller secret keys, and each of these
can be used independently of one another.  (By ``independently'',
we mean that one can encode a plain-text using the first key prior
to knowing the identity of the plain-text that will be encoded
using the second key).  This feature does not hold when
implementing private communication using a private SRF.

Although a private SRF is not equivalent to secret classical key or
entanglement, the former can yield the latter when supplemented by the
use of a public quantum channel. Specifically, one can distribute a
secret classical key by implementing the private classical
communication scheme outlined in this paper with the key as
plain-text.  Similarly, one can establish entanglement between two
parties by implementing a private quantum communication scheme where
the subsystem encoding the quantum plain-text is entangled with
systems that the sender keeps.  Note that a private SRF also yields
secret classical key if it is supplemented by a public SRF. For
instance, perfect private and public shared Cartesian frames yield an
infinite amount of secret key (in practice, the size of the key is
limited by the size of the physical system that defines the Cartesian
frame).

Another question of interest is how a private SRF is established.
Clearly, a public Cartesian frame together with an infinite classical
key yields a perfect private Cartesian frame (the key defines the
Euler angles of the private frame relative to the public frame).
Shared entanglement of a certain sort can also be consumed to align
local RFs~\cite{tez1,Aci01,Joz00}.  Another interesting possibility is
to set up the SRF by transmitting systems from Alice to Bob in a way
that is sensitive to eavesdropping. Whether an analogue of key
distribution can be achieved in this context is an interesting
question for future research.  Another such question is whether one
can recycle a private SRF by monitoring for eavesdropping, in the same
manner that one can recycle classical key and entanglement
\cite{Leu02,Opp03}.  Finally, we note that we have considered only
classical reference frames.  Preliminary research into the description
and characterization of \emph{quantum} reference frames
(c.f.,~\cite{Sch04,Col04}) leaves open the possibility for their use
as a shared private correlation.

Although the relationship between secret keys and entanglement has
been analyzed in some detail~\cite{Col01}, the relationship between
these and private SRFs still remains largely unexplored.  Quantifying
the power of private SRFs for encoding classical and quantum
information is an important step in such an investigation.

\textit{Note added in proof.} Recent independent results have
established the optimal schemes for transmitting an SU(2) reference
frame~\cite{Chi04,Bag04} and an $S_N$ reference ordering~\cite{Kor04}
through the transmission of quantum systems.  The techniques used in
these investigations are remarkably similar to those used to develop
our optimal private classical communication schemes using a private
shared RF.  Specifically, the optimal $N$-qubit states used for
transmitting a reference frame or reference ordering span many irreps
and are entangled between D-full/D-free subsystem pairs within an
irrep, as do the states used in our optimal private classical
communication schemes.

\begin{acknowledgments}
  T.R.\ is supported by the UK Engineering and Physical Sciences
  Research Council.  R.W.S.\ is supported by the Natural Sciences and
  Engineering Research Council of Canada. The authors gratefully
  acknowledge I. Devetak, D.\ Gottesman, D. Leung, M.\ A.\ Nielsen,
  and M.\ Plenio for helpful discussions.
\end{acknowledgments}


\begin{thebibliography}{99}

\bibitem {Ver26} G. S. Vernam, J. American Inst. Elec. Eng.,
  \textbf{55}, 109 (1926).

\bibitem {Ben92} C. H. Bennett and S. J. Wiesner, \prl \textbf{69},
  2881 (1992).

\bibitem {Amb00} A. Ambainis, M. Mosca, A. Tapp and R. de Wolf, in
  \textit{Proc. 41$^{st}$ Annual Symposium on Foundations of Computer
  Science}, (IEEE, Los Alamitos, 2000), p. 547.  Also
  arXiv:quant-ph/0003101.

\bibitem {Leu02} D. Leung, Quantum Info. Comp. \textbf{2}, 14 (2002).

\bibitem{Opp03} J. Oppenheim and M. Horodecki, arXiv:quant-ph/0306161.

\bibitem {BRS03a} S. D. Bartlett, T. Rudolph and R. W. Spekkens, \prl
  \textbf{91}, 027901 (2003).

\bibitem {Per02} A. Peres and P. F. Scudo, in \textit{Quantum Theory:
    Reconsideration of Foundations}, ed. A. Khrennikov (V\"axj\"o
  Univ. Press, V\"axj\"o, Sweden, 2002), arXiv:quant-ph/0201017.

\bibitem{GisPop} N. Gisin and S. Popescu,
  Phys. Rev. Lett. \textbf{83}, 432 (1999)

\bibitem {Per01} A. Peres and P. F. Scudo, \prl \textbf{86}, 4160
  (2001).

\bibitem {Bag01} E. Bagan, M. Baig, A. Brey, R. Mu\~noz-Tapia and R.
  Tarrach, \pra \textbf{63}, 052309 (2001).

\bibitem {Per01b} A. Peres and P. F. Scudo, \prl \textbf{87}, 167901 (2001).

\bibitem {Bag01b} E. Bagan, M. Baig and R. Mu\~noz-Tapia, \prl \textbf{87},
  257903 (2001).

\bibitem {Lin03} N. H. Lindner, A. Peres, and D. R. Terno, \pra
  \textbf{68}, 042308 (2003).

\bibitem {Fur98} A. Furusawa, J. L. S\o rensen, S. L. Braunstein, C.
  A. Fuchs, H. J. Kimble, and E. S. Polzik, Science \textbf{282}, 706
  (1998).

\bibitem{Bra99} S. L. Braunstein and H. J. Kimble,
  Phys. Rev. Lett. \textbf{80}, 869 (1999).

\bibitem {Rud01} T. Rudolph and B. C. Sanders, \prl \textbf{87},
  077903 (2001).

\bibitem {Enk02} S. J. van Enk and C. A. Fuchs, \prl
 \textbf{88}, 027902 (2002).

\bibitem {Wis02} H. M. Wiseman, J. Mod. Opt. \textbf{50}, 1797 (2003).

\bibitem {Wis03} H. M. Wiseman, Proceedings of SPIE Vol. \textbf{5111}
  Fluctuations and Noise in Photonics and Quantum Optics,
  Eds. D. Abbott, J. H. Shapiro, and Y. Yamamoto (SPIE, Bellingham,
  WA, 2003), pp 78-91.

\bibitem {Wis04} H. M. Wiseman, J. Opt. B: Quantum
  Semiclassical Opt. \textbf{6}, 5849 (2004).

\bibitem {San03} B. C. Sanders, S. D. Bartlett, T. Rudolph, and
  P. L. Knight, \pra \textbf{68}, 042329 (2003).

\bibitem {Bar03} S. D. Bartlett and H. M. Wiseman, \prl \textbf{91},
  097903 (2003).

\bibitem {Ver03} F. Verstraete and J. I. Cirac, \prl \textbf{91},
  010404 (2003).

\bibitem{KMP03} A. Kitaev, D. Mayers and J. Preskill, \pra
  \textbf{69}, 052326 (2004).



\bibitem{BRS03b} S.~D.~Bartlett, T. Rudolph, and R.~W.~Spekkens,
  \pra \textbf{70}, 032321 (2004).

\bibitem {Wer89} R. F. Werner, \pra \textbf{40}, 4277 (1989).


\bibitem {Zan97} P. Zanardi and M. Rasetti, \prl \textbf{79}, 3306
  (1997).

\bibitem {Mas95} S. Massar and S. Popescu, \prl \textbf{74}, 1259 (1995).

\bibitem {Zan03} P. Zanardi, D. A. Lidar and S. Lloyd,
  \prl \textbf{92}, 060402 (2004).

\bibitem {Zan01b} P. Zanardi, \prl \textbf{87}, 077901 (2001).

\bibitem {Kni00} E. Knill, R. Laflamme and L. Viola, \prl \textbf{84},
  2525 (2000).

\bibitem {Zan01} P. Zanardi, \pra \textbf{63}, 012301 (2001).

\bibitem {Ful91} W. Fulton and J. Harris, \textit{Representation
    theory: a first course}, (Springer-Verlag, Berlin, 1991).


\bibitem {tez1} T. Rudolph, arXiv:quant-ph/9902010.

\bibitem {Aci01} A. Acin, E. Jan\'e and G. Vidal, \pra \textbf{64},
  050302(R) (2001).

\bibitem {Joz00} R. Jozsa, D. S. Abrams, J. P. Dowling and C. P.
  Williams, \prl \textbf{85}, 2010 (2000); E. A. Burt, C. R. Ekstrom
  and T. B. Swanson, \prl \textbf{87}, 129801 (2001); R. Jozsa, D. S.
  Abrams, J. P. Dowling and C. P.  Williams, \prl \textbf{87}, 129802
  (2001).

\bibitem {Sch04} N. Schuch, F. Verstraete and J. I. Cirac, \prl
  \textbf{92}, 087904 (2004).

\bibitem {Col04} D. Collins and S. Popescu, arXiv:quant-ph/0401096.

\bibitem {Col01} D. Collins and S. Popescu, \pra \textbf{65}, 032321
  (2002).
  
\bibitem {Chi04} G. Chiribella, G. M. D'Ariano, P. Perinotti and M. F.
  Sacchi, arXiv:quant-ph/0405095.

\bibitem {Bag04} E. Bagan, M. Baig and R. Munoz-Tapia, \pra
  \textbf{71}, 030301(R) (2004).
  
\bibitem{Kor04} J. Von Korff and J. Kempe, arXiv:quant-ph/0405086.

\end{thebibliography}
\end{document}